\documentclass[11pt,twoside]{article}
\usepackage{./asp2014}
\usepackage[skip=0.5\baselineskip]{caption}

\aspSuppressVolSlug
\resetcounters

\begin{document}

\title{Deuteration in starless and protostellar cores}
\author{Rachel K. Friesen$^1$, Maria T. Beltr\'an$^2$, Paola Caselli$^3$, and Robin T. Garrod$^4$ \\
\affil{$^1$National Radio Astronomy Observatory, Charlottesville, VA, USA; \email{rfriesen@nrao.edu}}
\affil{$^2$INAF-Osservatorio Astrofisico di Arcetri, Largo E. Fermi 5, I-50125 Firenze, Italy; \email{mbeltran@arcetri.astro.it}}
\affil{$^3$Max-Planck-Institut f\"ur extraterrestrische Physik, Garching, Germany; \email{caselli@mpe.mpg.de}}
\affil{$^4$University of Virginia, Charlottesville, VA, USA; \email{rgarrod@virginia.edu}}
}

\paperauthor{Rachel K. Friesen}{rfriesen@nrao.edu}{0000-0001-7594-8128}{National Radio Astronomy Observatory}{}{Charlottesville}{VA}{22903}{USA}
\paperauthor{Maria T. Beltr\'an}{mbeltran@arcetri.astro.it}{0000-0003-3315-5626}{INAF-Osservatorio Astrofisico di Arcetri}{}{Firenze}{State/Province}{I-50125}{Italy}
\paperauthor{Robin T. Garrod}{rgarrod@virginia.edu}{0000-0001-7723-8955}{University of Virginia}{Department of Astronomy}{Charlottesville}{VA}{22903}{USA}
\paperauthor{Paola Caselli}{caselli@mpe.mpg.de}{0000-0003-1481-7911}{Max-Planck-Institut f\"ur extraterrestrische Physik}{}{Garching}{}{85748}{Germany}

\begin{abstract}

In dense starless and protostellar cores, the relative abundance of deuterated species to their non-deuterated counterparts can become orders of magnitude greater than in the local interstellar medium. This enhancement proceeds through multiple pathways in the gas phase and on dust grains, where the chemistry is strongly dependent on the physical conditions. In this Chapter, we discuss how sensitive, high resolution observations with the ngVLA of emission from deuterated molecules will trace both the dense gas structure and kinematics on the compact physical scales required to track the gravitational collapse of star-forming cores and the subsequent formation of young protostars and circumstellar accretion regions. Simultaneously, such observations will play a critical role in tracing the chemical history throughout the various phases of star and planet formation. Many low$-J$ transitions of key deuterated species, along with their undeuterated counterparts, lie within the 60-110~GHz frequency window, the lower end of which is largely unavailable with current facilities and instrumentation. The combination of sensitivity and angular resolution provided only by the ngVLA will enable unparalleled detailed studies of the physics and chemistry of the earliest stages of star formation.  

\end{abstract}

\section{Introduction}

In the local interstellar medium, the relative abundance of deuterium to hydrogen is $\sim 10^{-5}$ \citep{oliveira_2003}. In cold, dense molecular cores, however, observations have revealed abundances of deuterated molecules relative to their hydrogen-containing counterparts (i.e., DCO$^+$ and HCO$^+$; N$_2$H$^+$ and N$_2$D$^+$) that are orders of magnitude greater than the elemental abundance of D/H \citep{wilson_1973}. Furthermore, doubly- and triply-deuterated species have been detected toward both starless and protostellar cores \citep[i.e., NH$_3$ and ND$_2$H, ND$_3$;][]{roueff_2000,loinard_2001,vandertak_2002,lis_2002}. 

In cold molecular clouds where HD is the primary reservoir of deuterium, enhancement of deuterated species can occur through the exothermic chemical reaction

\begin{equation}
\mathrm{H}_3^+ + \mathrm{HD} \rightleftharpoons \mathrm{H}_2\mathrm{D}^+ + \mathrm{H}_2 + \Delta E
\end{equation}

\noindent where $\Delta E = 232$~K \citep{millar_1989}. At very low temperatures ($T \lesssim 20$~K) the forward reaction dominates due to the small energy barrier \citep{millar_1989,roberts_2003,walmsley_2004}. Other deuterated molecules, like N$_2$D$^+$, then form through ion-molecule reactions between H$_2$D$^+$ and species such as N$_2$. At the high densities characteristic of the interiors of highly-evolved starless cores ($n \gtrsim 10^5$\ cm$^{-3}$), CO and other neutral species begin to deplete from the gas phase through freeze-out onto dust grains. This reduces the destruction rate of the deuterated ions thus formed \citep{caselli_1999}, further enhancing their abundances relative to their non-deuterated counterparts. 

Deuterated molecules - particularly of N-bearing species, due to the depletion of C-bearing molecules from the gas phase - are thus selective tracers of the coldest, densest gas in molecular clouds and star-forming cores. In these cold regions, observations of the low-$J$ transitions of deuterated molecules and their non-deuterated counterparts are particularly important to determine independently their excitation conditions and column densities, and hence the deuterium fractionation as a function of radius across a core. In addition, many of these species are also predicted to distinguish well between disks and pseudodisks in magnetised models of core collapse \citep{hincelin_2016}, and are thus critical to distinguish rotationally-supported disks from collapsing pseudodisks, and to measure infall.

In warmer regions, such as the hot core or hot corino surrounding young protostars, several mechanisms are possible to enhance the deuterium fractionation of the gas. Detections of deuterated forms of methanol and formaldehyde suggest that deuterium fractionation may also occur on the surface of dust grains \citep{aikawa_2001,willacy_2007}. Observations of circumstellar disks and the Orion bar also suggest that enhanced deuterium fractions may also be produced in the gas phase with CH$_2$D$^+$ as the catalyst instead of H$_2$D$^+$ \citep{parise_2009,oberg_2012}. Observations of CH$_2$D$^+$ are critical to distinguish between these mechanisms, but have been extremely limited by telescope sensitivity and receiver availability, with only a single tentative detection of the $1_{01} - 0_{00}$ (para) feature at 278~GHz \citep{roueff_2013}. 

\begin{table}[t]
\caption{Prominent transitions of deuterated molecules between 60 and 115 GHz}
\centering
\renewcommand{\arraystretch}{1.2}
\begin{tabular}{lcc}
\hline
\textbf{Molecule} & \textbf{Transition} & \textbf{Rest Frequency} \\
 & & (GHz) \\
\hline
o-CH$_2$D$^+$ &	1$_{1,0}$ - 1$_{1,1}$	& 67.273 \\
p-NHD$_2$ & 1$_{1,1}$ - 1$_{0,1}$	& 67.842 \\
D$^{13}$CO$^+$ & 1 - 0 & 70.733 \\
DCO$^+$ & 	1 - 0	& 72.0393 \\
D$^{13}$CN & 1 - 0 & 71.175 \\
DCN	& 1 - 0	& 72.415 \\
CCD	& 1 - 0	& 72.108 \\
DN$^{13}$C & 1 - 0 & 73.368 \\
DNC	& 1 - 0	& 76.306 \\
N$_2$D$^+$	& 1 - 0	& 77.109 \\
HDO	& 1$_{1,0}$ - 1$_{1,1}$	& 80.6 \\
o-NH$_2$D	& 1$_{1,1}$ - 1$_{0,1}$	& 85.928 \\
p-NH$_2$D	& 1$_{1,1}$ - 1$_{0,1}$	& 110.15 \\
CH$_2$DOH	& \em{Multiple}	& 67 - 95 \\
\hline
\end{tabular}
\label{tab:dlines}
\end{table}

\section{Scientific Importance}

The scientific importance of observing low-level transitions of deuterated molecules with high resolution and sensitivity in dense starless and protostellar cores is twofold: first, to understand the chemical evolution from cloud to core to stellar system, and to shed light on the origin of chemical species in our own Solar System; and second, to better trace the physical structure and kinematics of dense cores on small, $\sim$~tens of au scales as they collapse under self-gravity to form protostars, circumstellar accretion regions, and protoplanetary disks. 

In dense cores, it is not yet clear what species are the best tracers of the cold molecular gas prior to the formation of a protostar. In some chemical models, all heavy elements are expected to deplete onto dust grains, leaving only H$_3^+$ and its deuterated isotopologues in the gas phase \citep{walmsley_2004}. Despite both H$_2$D$^+$ and D$_2$H$^+$ being observed toward multiple star-forming regions with single dish resolution \citep{stark_1999,caselli_2003,stark_2004,vastel_2004,friesen_2010}, H$_2$D$^+$ has only been detected with ALMA toward a single, low-mass prestellar core \citep{friesen_2014}. 

\begin{figure}
\includegraphics[width=\textwidth]{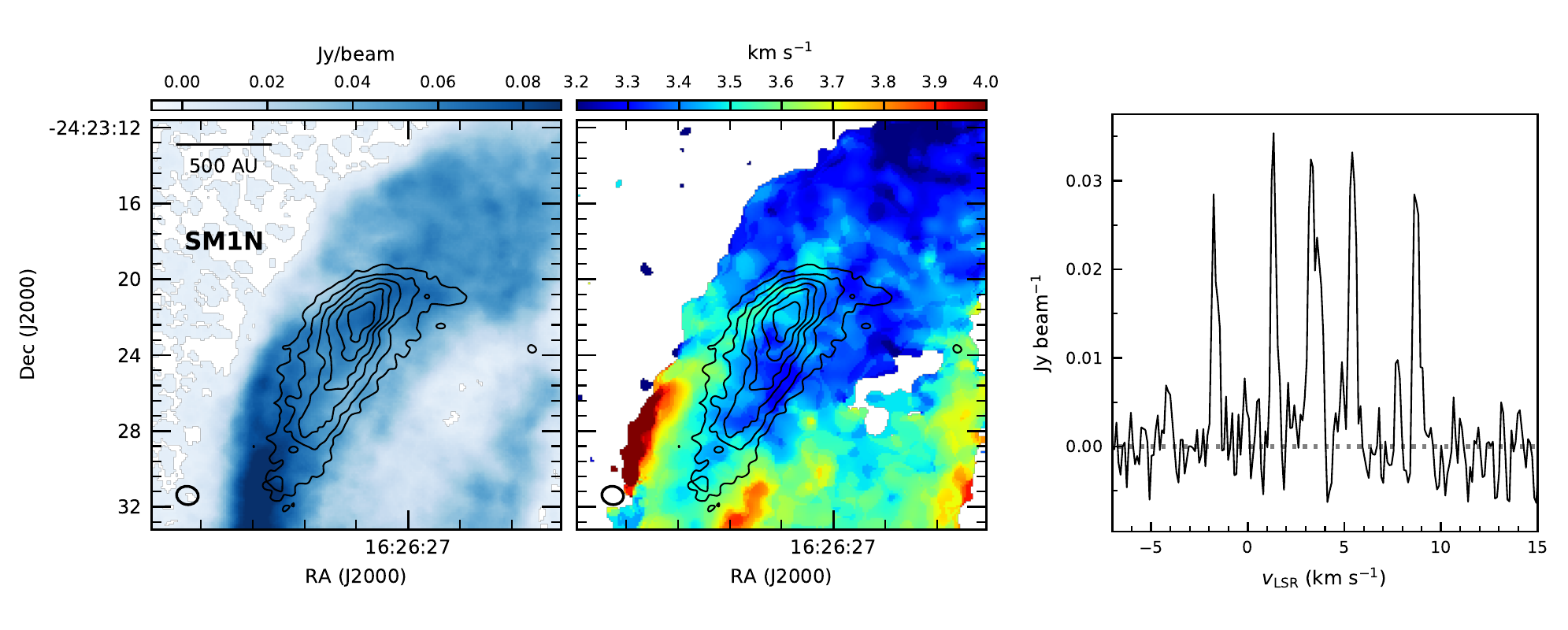}
\caption{NH$_2$D $1_{1,1} - 1_{0,1}$ integrated intensity (left), first moment (middle), and peak spectrum (right) observed with ALMA at 1\arcsec\ angular resolution toward the dense, star-forming SM1N core in the Ophiuchus molecular cloud (Friesen et al., in prep.). Black contours show 1~mm continuum emission at similar resolution. The NH$_2$D transition is just available at the lower edge of ALMA's Band 3. The hyperfine line strengths indicate that the emission is optically thick at the continuum peak, emphasizing the need to target these types of objects with denser gas-tracing deuterated species and rarer isotopologues. In addition, the extended nature of the potentially infalling gas around a young protostar requires both high resolution as well as a compact array component to recover larger-scale structures. \label{fig:alma_sm1n}}
\end{figure}

Detections of N$_2$D$^+$ and (singly- and doubly-) deuterated NH$_3$ toward evolved cores, however, imply that nitrogen-bearing species may remain abundant in these regions (see Figure \ref{fig:alma_sm1n}). In these cold regions, the ground-level rotational transitions of these species are the most sensitive probe of molecular abundances. Observations are needed to test and refine astrochemical models. The rest frequencies of these transitions for many key deuterated species lie below 80 GHz, however, and are not accessible by most current single-dish telescopes or interferometers (the IRAM 30~m Telescope and NOEMA interferometer can now observe at frequencies as low as 71~GHz, but with significantly lower resolution and sensitivity than will be available with the ngVLA). Some of these transitions are listed in Table 1. For many of these species, their non-deuterated counterparts also have observable transitions in this window, enabling direct measurements of the deuterium fractionation as a function of species. This band contains the low excitation lines of multiple deuterated species of interest in both starless cores and their more evolved, protostellar counterparts, observations of which are crucial to determining the relative importance of gas phase and grain surface reactions to the overall deuterium fractionation history of star-forming cores. In particular, this window contains the $J=1-0$ transitions of both N$_2$H$^+$ and N$_2$D$^+$, whose isolated hyperfine components enable high-precision measurements of the kinematics of the densest gas in star-forming cores. 


Once formed, young protostars heat their environments, forming `hot cores' (`hot corinos' in low mass objects) with sizes corresponding to where the heated gas reaches the sublimation temperature of the icy grain mantles ($\sim 100$ K). Many complex organic molecules are thought to form on warm dust grains ($\sim 30$ K - 40 K), and are returned to the gas phase within the hot core. Alternatively, some species may form in the gas phase from simpler molecules also released from the grains. Deuterated species are excellent tracers of whether a particular species forms in the gas phase or on dust grains. Measurements of the relative abundances of species like HCO$^+$, N$_2$H$^+$, NH$_3$, and CH$_3$OH and their deuterated forms in these environments are needed to identify the dominant deuterium fractionation routes in the gas-phase and on grains. These lines will provide detailed information on the region where ices are recently vaporised, testing both the kinematics and ice fractionation. Identification of deuterated methanol lines at higher frequencies in these objects is challenging due to the forest of emission lines present; at $60-80$\ GHz, however, the sparser line distribution allows the easy identification of several CH$_2$DOH lines within the band and an accurate measure of the deuterium fraction. Furthermore, only the ngVLA can enable sensitive, high resolution observations of the low energy transition of CH$_2$D$^+$ at 67~GHz to test directly its importance to deuterium fractionation in protostellar cores and disks. 

\section{Observing deuterated species with the ngVLA}

The ngVLA is best suited to detect and map the emission between 60 - 110~GHz from deuterated species at resolutions of a few tens of au toward dense molecular cores in nearby star-forming regions in the Galaxy. The ngVLA's sensitivity in this regime will enable detailed maps of the distribution of multiple deuterated species (simultaneously with their non-deuterated counterparts in some cases) across starless and protostellar cores with resolution down to only a few tens of au. With this sensitivity and resolution, we can trace the deuteration trail from starless cores to protostellar envelopes, and further to the evolution of protoplanetary disks. 

Evolved starless cores with large contrast between their central and mean density are the ideal targets for this study. Furthermore, cores on the cusp of star formation are the best targets to investigate collapse dynamics, since at earlier stages, the infall velocity has yet to reach the supersonic values predicted in hydrodynamic models, while at later times the collapse motions may be masked by strong outflows.

For these objects, sensitive imaging at $\sim 65$~GHz to 90~GHz with angular resolution $\lesssim 1$\arcsec\ will provide physical resolutions of $\sim 100$~au$ - 500$~au in the nearest star-forming regions. This resolution is needed to resolve the peak of the density distribution of typical evolved starless cores, to study the chemical composition of the innermost regions (where stellar systems will form), and test current theories of complete freeze-out of elements heavier than He. Within the cores, the turbulent motions of the gas are largely dissipated, and the lines are thus expected to be narrow. Kinematics on these size scales are critical to distinguish between rotationally supported disks and pseudodisks at young ages (Harsono et al. 2015), and to track the increase in infall velocity at small scales predicted by hydrodynamic models (e.g., Foster \& Chevalier 1993). 

\begin{figure}
\includegraphics[trim=0 270 5 280,clip,width=\textwidth]{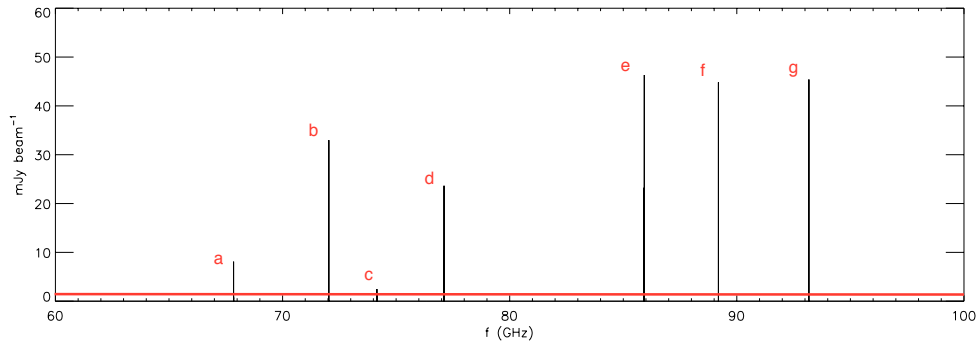}
\caption{Predicted line brightnesses in a 1\arcsec\ beam toward an evolved, starless core based on a chemical model by \citet{harju_2017}. Lines include a) p-NHD$_2$ ($1_{1,1} - 1_{0,1}$), b) DCO$^+$ ($1-0$), c) NH$_2$D ($2_{1,2} - 2_{0,2}$), d) N$_2$D$^+$ ($1-0$), e) o-NH$_2$D ($1_{1,1} - 1_{0,1}$), f) HCO$^+$ ($1-0$), g) N$_2$H$^+$ (1-0). The red line shows $\sim 2$~mJy~beam$^{-1}$. \label{fig:harju}}
\end{figure}

We have calculated the line strengths for a number of deuterated species and their non-deuterated counterparts in a starless core based on density, temperature and abundance models derived for the Ophiuchus H-MM1 core by \citet{harju_2017}, assuming 0.25 km/s velocity dispersion, 1\arcsec\ angular resolution, and a distance of 250~pc. Figure \ref{fig:harju} shows the resulting line brightnesses of NH$_2$D, N$_2$H$^+$, N$_2$D$^+$, DCO$^+$ and HCO$^+$ transitions from Table \ref{tab:dlines}. Under the assumption that the deuterium fractionation levels are close to those currently measured toward starless cores, a sensitivity of 2~mJy~beam$^{-1}$ will provide good S/N detections for most species, as well as for the hyperfine structure of the NH$_2$D lines, enabling precise measurements of line excitation and opacity needed for abundance analysis. We therefore require a sensitivity of 2~mJy~beam$^{-1}$ (0.4~K) within a velocity resolution of 0.1~km~s$^{-1}$. Ideally, a flexible spectral setup will allow the targeting of multiple lines within a single observation. 

In protostellar cores, higher angular resolution is needed to resolve the hot core or hot corino, where the bulk of the deuterated organic species are expected to be found, and to determine the critical processes driving the enhanced deuterium fractionation in these environments. The physical hot core/corino radii range from several tens of au for low-mass sources to $\sim 1000$~au for high mass sources. An angular resolution of 0.2\arcsec\ will match the predicted corino size for nearby sources (spatial resolution of 50~au at a distance of 250~pc), and resolve the hot core in more distant, high mass sources (spatial resolution of 400 au at a distance of 2~kpc). Lines will be broader than in the cold cores, allowing a larger velocity resolution of 1.5~km~s$^{-1}$. Simulations of CH$_2$DOH in a 1\arcsec\ beam toward a typical massive hot core suggest that a brightness sensitivity of 0.3~K will allow the detection of multiple lines within the band \citep{beltran_2015}. In a 0.2\arcsec\ beam, the equivalent flux density sensitivity thus required is 70~$\mu$Jy~beam$^{-1}$. Spatial differentiation between species such as N$_2$H$^+$ and N$_2$D$^+$, and variations in the deuterium fractionation of N$_2$D$^+$ between sources, further test chemical evolution models \citep{tobin_2013} and may provide an elusive age indicator for the youngest protostars \citep{emprechtinger_2009,fontani_2011}. 

At the very small physical scales probed here, we are targeting very high density regimes. Some lines may be optically thick in starless cores, while dust opacities may be high in protostellar cores. Rarer isotopologues of the targeted species are present within the proposed frequency band (see Table 1) and, in case dust opacity is an issue, heavier and more complex deuterated molecules at lower frequencies may also be targeted.

\section{Uniqueness to ngVLA Capabilities}

Historically, very few telescopes (neither interferometer nor single dish) have had detectors working in this frequency band, leading to a major dearth of data on these low-level transitions of important deuterium-bearing species in star-forming cores. Ideally, a frequency range of $\sim60$-110~GHz would allow measurements of the critical species discussed here. In particular, no other telescope can observe both the N$_2$D$^+$ and N$_2$H$^+$ $J=1-0$ lines at 77~GHz and 93~GHz, respectively, with the sensivity and resolution required for deuterium studies in starless and protostellar cores. As of 2017, an upgrade of the NOEMA interferometer allows observations between 71 GHz and 80~GHz, but with orders of magnitude less sensitivity and imaging capability as the ngVLA.  Furthermore, the frequency band between 60 GHz and 71 GHz is not covered in whole by any other telescope at this time, such that the 67~GHz CH$_2$D$^+$ line is currently unobservable with current facilities. 

In the future, Band 2 and/or Band 2/3 receivers will likely be developed
for ALMA that cover the low-level transitions of deuterated species
discussed here. ALMA observations, however, will lack the needed
sensitivity to map these species toward dense cores at resolutions of
a few tens of au in a typical, nearby star-forming cloud. No other
planned facilities combine the required sensitivity and resolution of
the ngVLA. 

The sensitivity and high angular resolution of the ngVLA are
critical to mapping the distribution of deuterated species across
starless and protostellar cores. In particular, observations with the
ngVLA will be able to target, for the first time, the chemical
transition at only a few tens of au between hot core and cold
envelope, where we expect a transition between the dominant chemical
pathways to enhanced deuterium fractionation. 

Infrared absorption observations with JWST and ELTs targeting young
protostars will enable detailed analysis of the ice composition on
dust grains, providing complementary information to the gas-phase
observations enabled by the ngVLA. In combination, this will allow
rigorous tests of astrochemical models of grain-surface and gas-phase
deuterium fractionation, crucial to exploit fully the diagnostic power of deuterated molecules and their chemical history throughout the various phases of star and planet formation.

\acknowledgements The National Radio Astronomy Observatory is a facility of the National Science Foundation operated under cooperative agreement by Associated Universities, Inc.  

\bibliography{biblio}  



\end{document}